# The Development of Gaseous Detectors with Solid Photocathodes for Low Temperature

L. Periale[1,2], V. Peskov[3], C. Iacobaeus[4], T. Francke[5], B. Lund-Jensen[3], N. Pavlopoulos[1,6], P. Picchi[1,2], F. Pietropaolo[1,7]

[1] CERN, Geneva, Switzerland
[2] IFSI-CNR of Torino, Torino, Italy
[3] Royal Institute of Technology, Stockholm, Sweden
[4] Karolinska Institute, Stockholm, Sweden
[5] XCounter AB, Danderyd, Sweden
[6] Leonardo deVinchi Universiy, Paris
[6] INFN Padova, Italy

**Abstract**

There are several applications and fundamental research areas which require the detection of VUV light at cryogenic temperatures. For these applications we have developed and successfully tested special designs of gaseous detectors with solid photocathodes able to operate at low temperatures: sealed gaseous detectors with $MgF_2$ windows and windowless detectors. We have experimentally demonstrated, that both primary and secondary (due to the avalanche multiplication inside liquids) scintillation lights could be recorded by photosensitive gaseous detectors. The results of this work may allow one to significantly improve the operation of some noble liquid gas TPCs.

## I. Introduction

Cryogenic liquid gas TPCs are promising detectors for several applications and fundamental researches. Relevant examples could be the ICARUS experiment [1], the nTOF experiment [2], WIMP search LXe/Ar detectors [3], noble liquid gas PETs [4], studies of cryogenic plasmas [5] and studies of quantum phenomena in liquid and solid He [5]. Nowadays, PMs are mostly used for the detection of the primary scintillation light from noble gas liquids (a "start" signal for the TPC). In preliminary studies we have shown that the costly PMs could be replaced by gaseous detectors with solid photocathodes placed far away from the volume with noble gas liquids and operating at room temperature [6]. Recently, we have developed and successfully tested several exclusive designs of gaseous detectors which are able to operate at cryogenic temperatures. Their main advantages are their low cost, the large sensitive area and their ability to operate in magnetic fields, which is quite necessary for some studies. In contrast to the previously used gaseous detectors, the new ones could be placed much closer to the liquids or if necessary, be even immersed into the liquids. The aim of this report is to review these new but yet unpublished results.

## II. Experimental set up

Two experimental set ups were used in this work: one oriented for work with cooled gases and the other one for measurements in vacuum, gases and tests with noble gas liquids.

The firsts experimental set up is schematically presented in Fig.1. It basically consists of a cryostat, inside of which a test vessel was installed. The cryostat could be cooled in a controllable way down to 80K. The test vessel was comprised of a gas scintillation chamber filled with noble gases (Ar, Xe or Kr) and contained a radioactive source ($^{241}$Am, $^{55}$Fe or $^{90}$Sr) inside, a gaseous detector with solid photocathode attached to the scintillation chamber and the PM (EMI 9426 with a MgF$_2$ window) monitoring the primary scintillation light produced by the radioactive sources.

Two types of photosensitive defectors were constructed, manufactured and tested: 1) sealed detectors with MgF$_2$ windows and 2) windowless detectors able to operate in cooled noble gases (see [7,8] for more details). In the case of the sealed detectors, most of the measurements were done with a single wire counter (a diameter of the anode wire was 50 μm) shown in Fig.2. Inside the cylindrical cathode (35 mm in diameter) was a removable disc (30 mm in diameter) coated with a photosensitive layer. If necessary, this disc could be independently cooled (see [7] for more details).

Photosensitive layers were deposited either on the surface of the disc (reflective photocathode) or on the MgF$_2$ widow (semitransparent photocathode). Most of the tests were done with reflective CsI, Sm, TMAE and TMAE+neopentane (NP) photocathodes. The procedure of deposition of these layers on the cathode's disc is described in [7]. Typically, the thickness of the semitransparent CsI layer was 20 nm and the reflective one – 400nm. The thicknesses of the other photosensitive layers was, depending on the conditions, between 0.5 and 2 μm. Sealed detectors were filled, depending on the operation's temperature, either with a mixture of Ar+10% CH$_4$ or He +(5-10)% H$_2$ at pressures of 1atm. For the estimation of the number of detected scintillation photons N$_{ph}$, a $^{55}$Fe source was often used which produces a known number of primary electrons (see [6] for more details). For the independent measurements of the photocahode's quantum efficiency (QE) at room temperature a calibrated gaseous detector CFM- 3 was used [9].

Some measurements were done with Hamamatsu capillary plate (CPs) -see Fig 3. The conventional CPs were made of lead glass and had a resistivity between the electrodes of > 10$^{10}$ Ω. We also tested H$_2$ treated CPs which had the resistivity between the electrodes of 10-100 MΩ. Both types of CPs had a diameter of 25mm, thickness of 0.8 mm and the hole's diameter of 100 μm In tests described below the CPs were not coated with any photosensitive layer.

Windowless detectors tested were either a cascade of hole-type detectors (GEM or capillary plates[10]) with their (see Fig. 4) or a parallel mesh detector combined with a reflective CsI photocathode (Fig.5). The hole-type detector operated in an avalanche multiplication mode in the same gas and pressure as the scintillation chamber. In the case of GEM its cathode facing the Am source coated with a 0.4 µm thick CsI layer. As was shown earlier [6], in this mode of operation the QE is low compared to the detectors with windows. However, the sensitive area of these detectors could be rather large and thus the "overall" sensitivity could be reasonably high. This is why it was still interesting to investigate windowless detectors at low temperature. In the case of the mesh detector, we tried to explore, not the charge, but a multistep light multiplication. The detector works as follows. Photons from the primary scintillation create $n_0$ primary electrons from the CsI cathode. These electrons drift to the space between the two meshes and produce $n_0 A_{ph}$ secondary scintillation photons. These photons in turn create a well separated bunch of secondary electrons in time from the cathode, $n_1 = n_0 A_{ph} \Omega Q$ (where $\Omega$ is the solid angle, Q-the photocathode QE) and so on (see [7,8]). As a result of two/three generations in these processes, the charge signal becomes large enough to be detected. This approach has several important advantages over other known methods and we consider it as the most promising one for this application.

The other set up was a chamber which could be immersed to the bath cooled with $LN_2$ or other liquids. And if necessary be filled with noble gas liquids as described in [7, 11] –see Fig.6. It allows several independent studies to be made, for example: operation of hole-type structures placed above the liquid's level, the avalanche multiplication inside the liquids, detection of the primary and the secondary scintillation lights by a PM or by a gaseous detector with solid photocathodes. In addition, it was possible to measure the QE of these photocathodes both in the vacuum and in a gas at some temperature intervals, including those, which corresponded to LXe or LAr. For this, a pulsed (a few ns) $H_2$ or a continues Hg lamp was used with a system of UV filters. The absolute intensity of the light beam was measured with calibrated Hamamatsu vacuum photodiodes and the calibrated SFM-3 counter (see above and Fig.7). Some measurements were also done in the visible region of spectrum.

### III. Results
### III-1. Resulst Obtained with the First Set Up
*A) Detectors with windows*

The absolute values of the photocathode's QE were estimated from measurements made at room temperature by three methods: 1) from the measurements of the amplitude of the signal produced by the scintillation light [6,7] (at very low QE the counting rate produced by single photoelectrons was used instead of the signal amplitude), 2) with respect to the known QE of the SFM-3 counter (see Fig.7 and 3) with respect to the known QE of (ethylferrocene) EF[12] and TMAE vapours (using the scintillation light from the noble gases). In the latest case the detector was filled with TMAE or EF [12] and the amplitude of the signal produced by alpha's scintillation was measured with respect to the $^{55}$Fe signal. Some measurements were done with the pulsed $H_2$ lamp. The results are presented in Table-1 Note that $CH_4$ mixture absorbs the Ar scintillation light and thus a special calibration was used to correct on this absorption [7]

At low temperatures we indirectly monitored the QE by measuring only the scintillation signal produced by the $^{241}$Am. During cooling all valves were closed, so the gas density was constant. Fig. 8 shows the signal amplitude vs. temperature for the CsI, TMAE, Sm, and NP+TMAE reflective photocathodes (in VUV and visible regions of spectra). In these measurements special care was done to minimize the temperature gradient inside the detector which may cause some undesirable gain variations. At temperatures of T>120K the detector was filled with Ar+$CH_4$ gas mixture. In this gas the detector operated stably at gains of $10^4$-$10^5$. At T<120K the gas mixture was He+$H_2$. Our measurements show that in this mixture due to the strong back diffusion effect [13] the "practical" QE drops, depending on the gas density, by a factor of 5-10 (compared to the Ar+CH$_4$ mixture) and this in turn caused a drop in the signal amplitude to take place. This effect is clearly seen from Fig. 8. The other problem associated with the He+H$_2$ mixture was a rather narrow proportional region at an elevated gas density, for example at P=1atm and at T=80K: the detector could at gains of only 200-500 easily transit to the streamer mode of operation. Obviously this was due to the large diameter of the anode wire (see [14] for more details).Also note that for the stable operation at T<100K, the inner part of the detector and the gas should be very clean, otherwise the impurities could condense on the photocathode's surfaces and created some charging up effects. This also concerns the CsI photocathode: it should not be exposed to air and should also be well outgased; otherwise some charging up effect is possible.

The other interesting observation was that in VUV even pure metals (for example, Sm or Cu) may have high QE, however in UV and visible region their QE drop quickly with temperature. The high QE of metals in the VUV region of spectra allows us to use CPs for detection of the scintillation light without any coating by photosensitive layers. These test reveal that conventional CPs could be used without any serious problems up to 200 K and H$_2$ treated up to 150K.

*B) Windowless detectors*

As in the previous case, alpha particles were used for QE monitoring. With windowless detectors by changing the polarity we were able to measure either the signal produced by the scintillation light or the charge signal produced by alpha particles in the gas. At high gains $^{55}$Fe was also used for signal's calibration. We discovered that GEMs can work in pure noble gases at T>120 K (see also [15]). However, at lower temperatures a strong charging up effect was observed. We also discovered that at room temperature much higher gains than with GEMs were achieved with capillary plates (CPs) operating in pure noble gases. For example gains up to $10^4$ were easily achieved with single CPs as shown in Fig.9 (see [7] for more details). This is why we mostly concentrated on studies of CPs operations at low temperatures. However, even the H$_2$ treated CPs exhibited charging up problems below 120K. The other problem, common both for GEMs and CPs, was that the QE was low compared to the detectors with windows [6]. This can also be clearly seen from Fig. 8 in which the plotted signal produced by the $^{241}$Am scintillation light is seen.

By almost a factor of 5 times higher the QE was obtained with the mesh detector– see Fig.8 and [7] for more details. The other advantage of this detector was that no charging effect was observed at any temperature. However, more studies still are needed to prove the practical significance of this method.

### III.-2 Results with the Second Set Up

As was mentioned above, the second set up (see Fig.6) allows one to carry out various studies including measurements with noble gas liquids. For example recently, photosensitive gaseous detectors were successfully used in the detection of both the primary and the secondary (due to the avalanche multiplication inside the noble liquids) scintillation lights [7,8]see Fig 6. This work was in the stream of other efforts described in recent papers [11]. In this report however we will present only the results of the QE measurements in vacuum and in the gas at very low temperature (up to LN$_2$) and mention some studies of the detector's operation in an avalanche mode. The description of other results can be found in [7]. In Table 2 the results of the QE measurements are presented for a single wire detector installed inside the test chamber. The drop of the QE at 89K in He+H$_2$ gas was, as mentioned above, due to the back diffusion. Note that this detector could operate stable at high gains and at any temperature, the only bothersome effect was the transition to the streamer's mode of operation at temperatures of T>100K. In contrast, brief tests with the hole type detectors revealed some instability below 150K, probably due to the charging up effect or some contacts problems. For example, in a recent attempt to operate GEMs placed few cm above LAr surface were not successful. More studies are necessary to clarify the reasons for the detectors failure. .

### III.-3. Estimation of the QE of detectors operating at low temperatures

From the data presented in Fig. 8 and Tables 1,2, and using the results of the sensitivity calibrations, one can estimate the QE of our detectors (see [7] for more details). Results are presented in Fig 10. One can see that the highest QE was achieved with the sealed single wire counter. The parallel mesh detector or GEMs coated with the CsI layer offer lower efficiency, however they can operate without window which could be attractive in some applications. Thus photosensitive gaseous detectors could operate at low temperature with a high enough QE.

### IV. Discussion and Conclusion

We have demonstrated that gaseous detectors with solid photocathode can operate stable, depending on designs, up to 150-80K.
The best results (the highest QE, the highest gains and good stability) were obtained with sealed gaseous detectors operating under very clean conditions.
This confirmed our earlier results obtained with solid photocathodes operating inside liquid and solid noble gases [16]. It is especially important that they have the ability to operate in magnetic fields. As was described in other papers, one can also explore avalanche multiplication inside noble liquids. All these results may allow one to significantly improve the operation and sensitivity of the TPCs and reduce their cost. Some designed proposed on the basis of these studies are described in [7].

**VI. Figures and tables:**

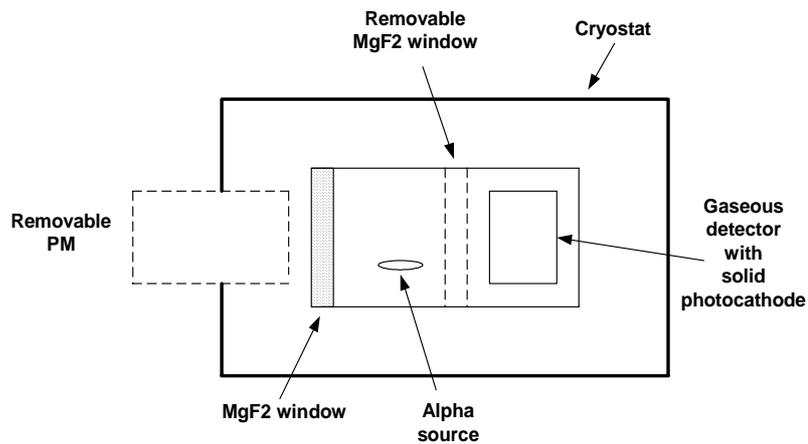

Fig.1 A schematic drawing of the first experimental set-up.

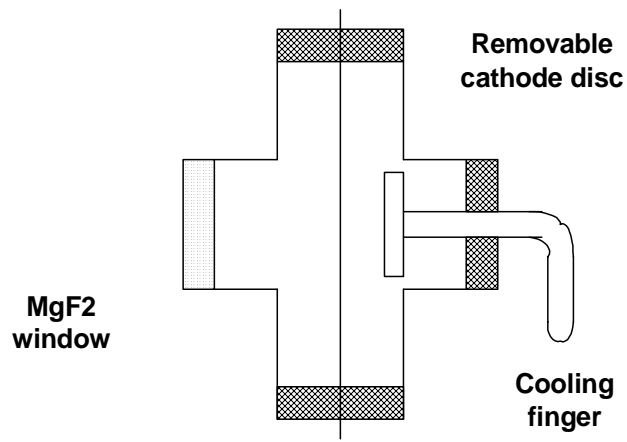

Fig.2 One of the designs of a single wire counter used in our test.

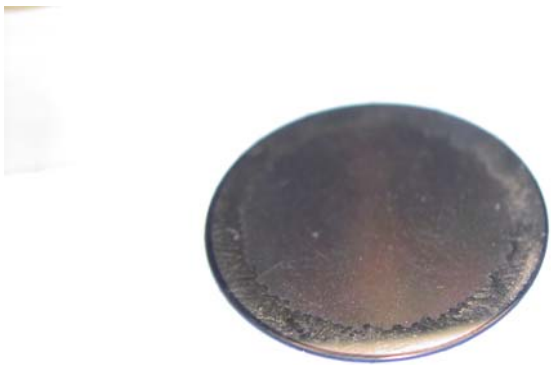

Hamamatsu capillary plate.

Fig. 3 The Hamamatsu capillary plate.

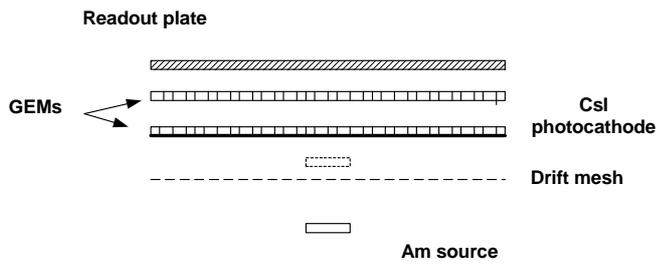

Fig. 4 A schematic drawing of cascaded GEMs used in our tests.

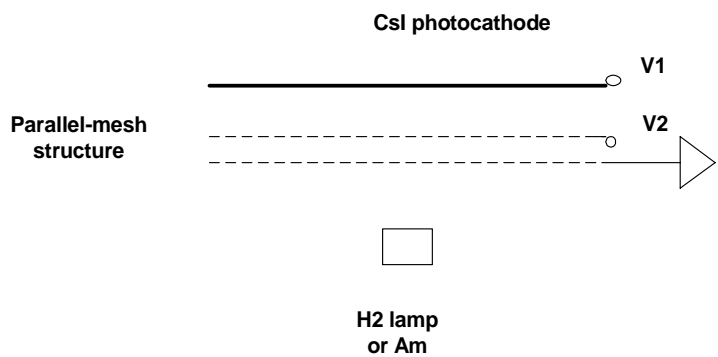

Fig.5. A schematic drawing of the parallel-mesh detector.

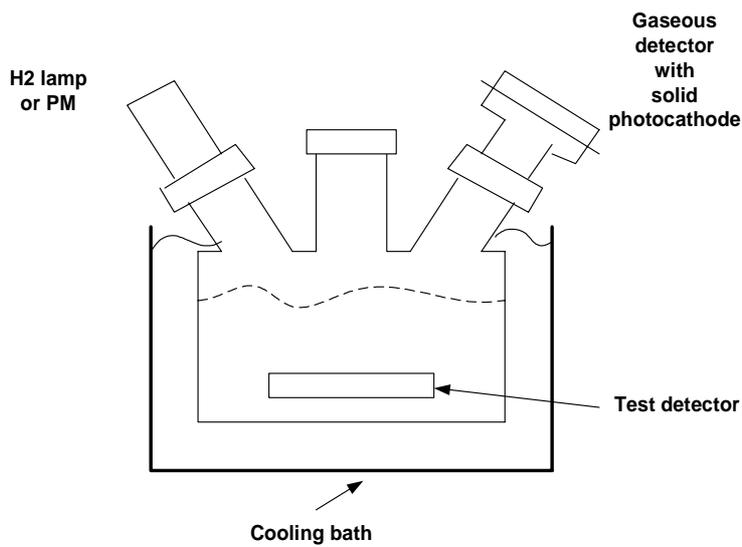

Fig.6 A schematic drawing of the second set up.

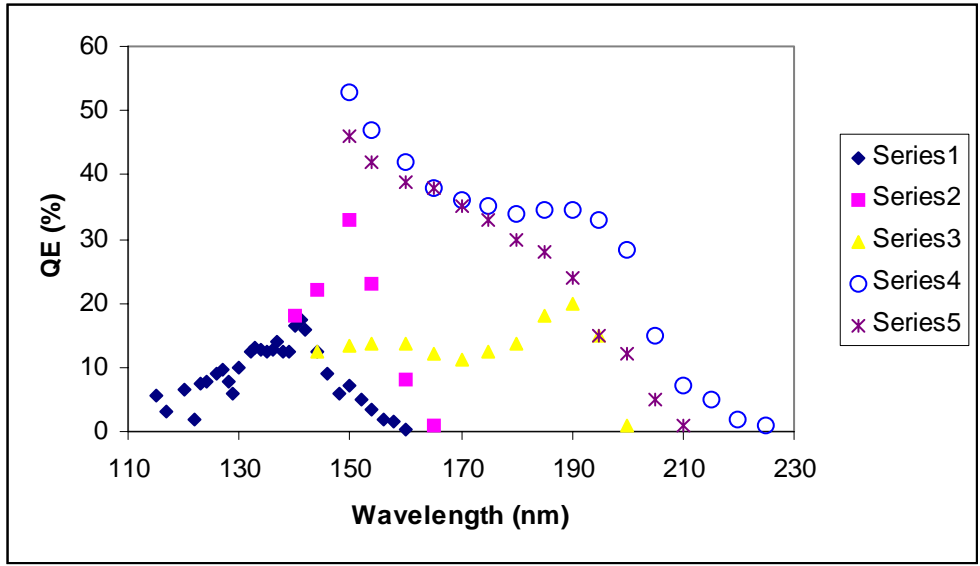

Fig. 7 The quantum efficiency (QE) of some photosensitive materials: 1-vapours used in SFM-3 counter, 2-TEA vapours, 3-EF vapours, 4-TMAE vapours. 5- For comparison the QE of CsI photocathode measured in vacuum is given.

| Photocathode / Detector Type | CsI at $\lambda$=120-130 nm (Ar Light) | CsI at $\lambda$=175 nm (Xe Light) | CsI at $\lambda$=165 nm ($H_2$ lamp) | TMAE at $\lambda$=120-130 nm (Ar Light) | TMAE+NP at $\lambda$=120-130 nm (Ar Light) |
|---|---|---|---|---|---|
| Single wire (Ar+$CH_4$) | 19 (1) 37,4 (2) | 18,3 (1) 26 (2) | 23,5 (3) | 0,4 (1) | 0,6 (1) |
| Single Wire in (He+$H_2$) | 2,3(1) 3.1(2) | 2 (1) | 3,2(3) | 0,15 (1) | 0,25 (1) |
| GEM (Ar) | 6(1) | | 4(3) | | |
| PPAC (Ar) | 13 (1) | | 7,8(3) | | |

Table 1
QE (%) of various gaseous detectors, measured with the first set up at room temperature (notes in bracket indicate the method of measurements):1-amplitude of signal produced by $^{241}$Am in Ar or Xe, 2-with respect to the QE of EF, 3-with respect to CFM-3 counter). In the case of the single wire counter (SWC) filled with Ar+$CH_4$ mixture and detecting the Ar scintillation light the gas gain was ~$10^4$. In the case of the parallel-mesh detector the gain was ~30. In all other measurements the gain was ¨~100.

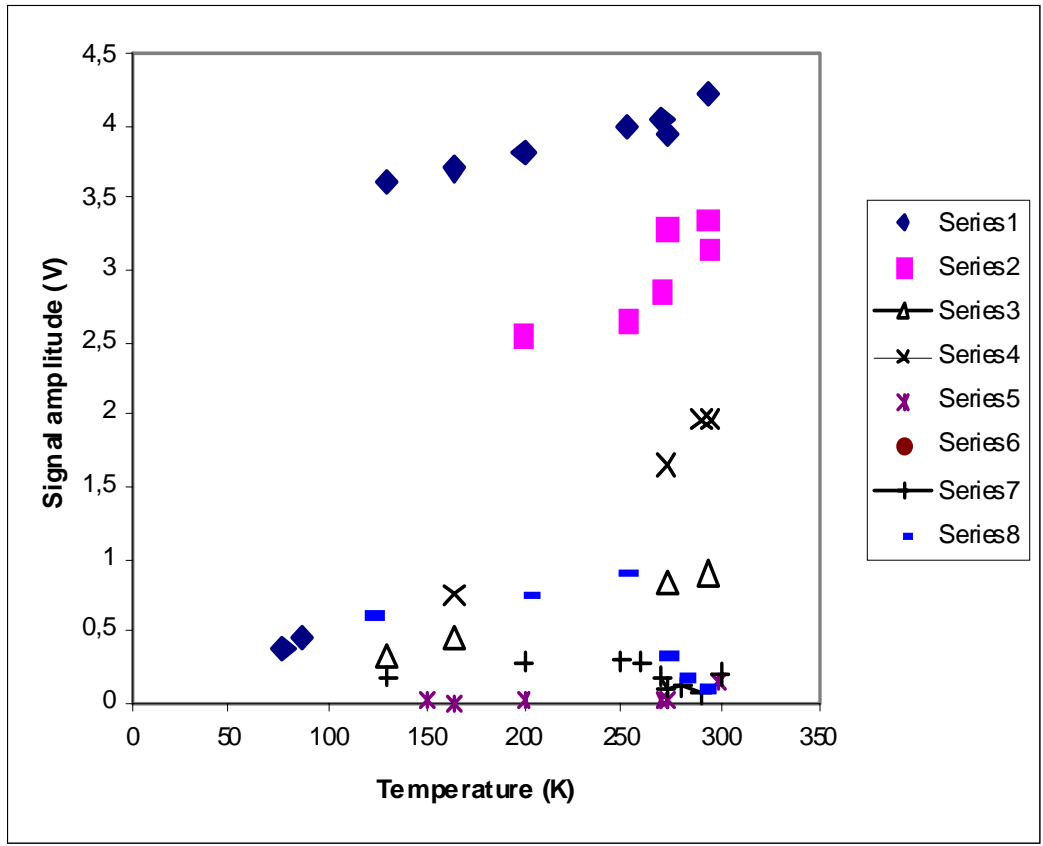

Fig. 8. Measured (and recalculated from the experimental data to normalise for the same gain) signal amplitudes produced by the $^{241}$Am scintillation light for various photocathodes and detector's designs at a gain of 100.
1- a single wire counter (SWC) with CsI photocathode filled with Ar+CH$_4$ (for T>130K) and He+H$_2$ for T<130K) and detecting Ar scintillation light, 2- the same detector filled with Ar+CH$_4$ and detecting Xe scintillation light, 3-GEM detecting Ar scintillation light, 4- parallel-mesh structure detecting Ar scintillation light, 5-open circle- SWC with a Sm photocathode detecting Ar scintillation light, 6-the same, but vissible light, 7- SWC with a TMAE photocathode, 8-SWC with a TMAE+NP photocathode

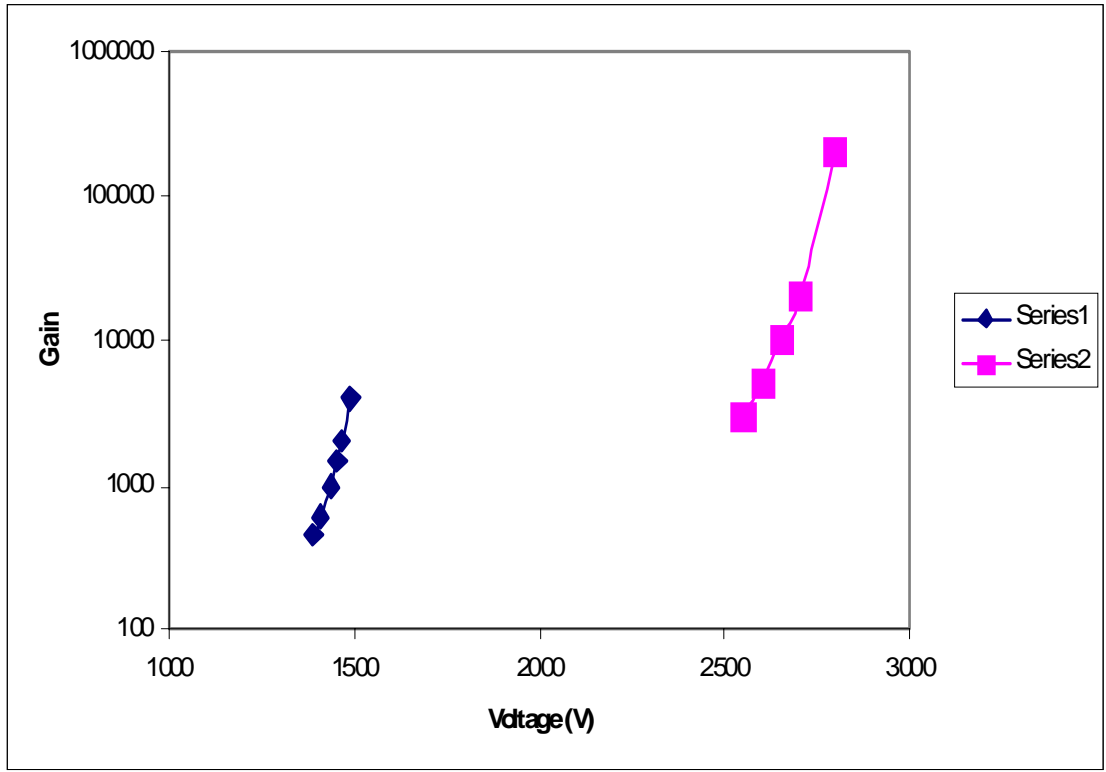

Fig.9. Gain vs. voltage for a single CP (1) and triple GEMs (3).

| T(K) / Detectors type | 77 | 88 | 130 | 293 |
|---|---|---|---|---|
| Single wire (Ar+CH$_4$) | | | 21 | 23.5 |
| Single wire He+H$_2$ | 1,1 | 1,5 | 2,2 | 2,5 |
| Sihgle wire (vacuum) | 26 | 27 | 29 | 32,6 |
| GEM (Ar) | | | 2,5 | 8 |

Table2. QE (at 165 nm) of CsI photocathode measured with the second set up at some selected temperatures.

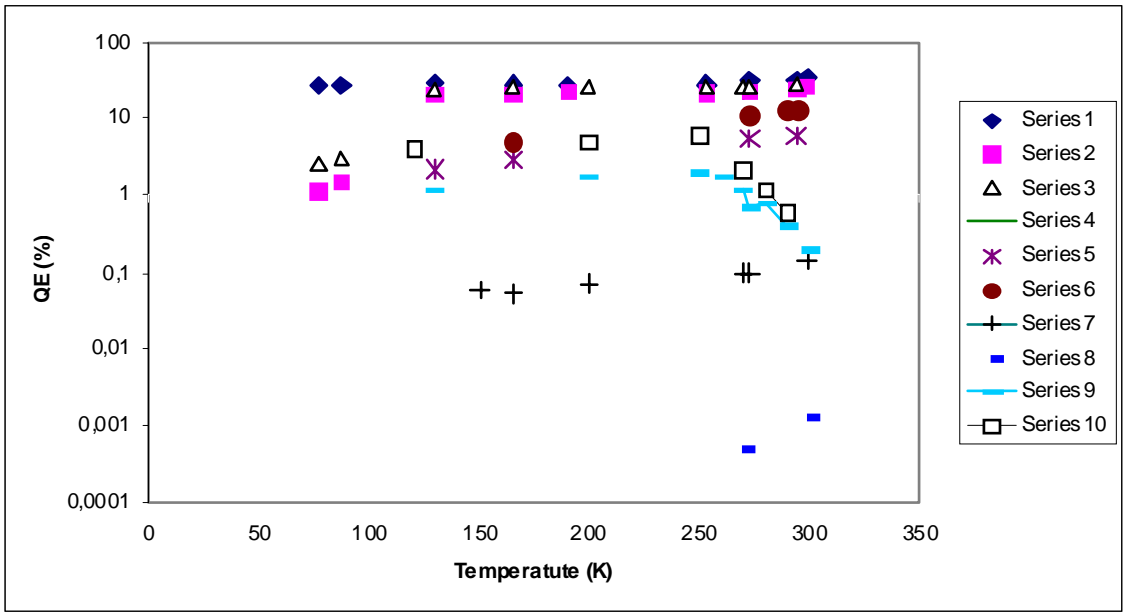

Fig. 10 Calculated QE (for all results combined together and normalized to gain of 100): 1-vacuum, $H_2$ lamp; 2-single-wire, $H_2$ lamp; 3-Single wire, Ar light; 4-the same, Xe light; 5-GEM, Ar light; 6-parallel-mesh, Ar light; 7-Sm, Ar light; 8- Sm, visible light; 9-TMAE, Ar light; 10-TMAE+NP, Ar light.
.